# Social sycophancy differs across access modes of large language models

Alek Klem and Uwe Peters*
Utrecht University

*Abstract*—**Large language models (LLMs) can display sycophancy, a tendency to excessively praise, agree with, or validate a user. Existing studies have tested LLMs for sycophancy primarily through application programming interfaces (APIs) rather than the consumer user interfaces (UIs) through which the public commonly interacts with LLMs. Access mode (API vs. UI) may affect observed sycophancy rates. However, it remains unknown to what extent prominent models show API–UI differences in social sycophancy, which includes behaviors potentially more likely in UIs (e.g., softening warranted criticism). We tested ChatGPT, Claude, Gemini, and DeepSeek through both APIs and UIs on four dimensions of social sycophancy. Across models, UI access was generally associated with greater social sycophancy, with odds ratios up to 7.56 (95% CI [6.22, 9.20]). However, UI–API differences varied significantly across models for all sycophancy dimensions, including a pronounced reversal for DeepSeek moral sycophancy (43% UI vs. 82.3% API). Overall, our results suggest that API-based LLM evaluations may underestimate the prevalence of sycophancy in corresponding consumer-facing systems, making access conditions consequential for the external validity of LLM evaluation. To facilitate research on access-mode effects, we release 57,000 model responses and *SycoUI*, a browser-automation pipeline for standardized cross-interface LLM evaluation.**

## I. INTRODUCTION

Artificial intelligence (AI) chatbots driven by large language models (LLMs), such as ChatGPT, are now widely used to find information, obtain advice, solve problems, or perform writing tasks [1]. But despite their usefulness, LLMs can also exhibit undesirable behaviors [2], [3]. One such behavior is sycophancy, the tendency of an LLM to excessively praise, agree with, or validate a user's claims, viewpoints, or actions, even when they are inaccurate or inappropriate [4], [5], [6]. Sycophantic models may reinforce inaccurate views in users, promote problematic actions, or exacerbate mental health problems [7], [8].

Evidence of sycophancy has been documented across a range of LLMs and tasks (e.g., [4], [7], [9]–[11]), leading some commentators to draw the broad conclusion that "AI chatbots are sycophants" [12]. However, LLMs may respond differently depending on how they are accessed. LLM providers (e.g., OpenAI) typically make their models available through two main access modes: *application programming interfaces* (APIs), which are primarily designed for developers and researchers, and *consumer facing user interfaces* (UIs), which are designed for and primarily used by lay users (e.g., the web-based interfaces of ChatGPT or Claude) [13], [14]. Compared to APIs, UIs often incorporate additional system instructions, safety mechanisms, routing decisions, or memory features that may result in LLMs producing different responses to the same prompt across APIs and UIs [15].

Assessing whether LLMs show systematic differences in sycophancy across UIs and APIs is important, as the findings may have significant societal implications. Since using API access requires more technical expertise, access mode differences may create social disparities, with technically skilled users potentially receiving less sycophantic responses than lay users relying on UIs [16]. Furthermore, findings of studies testing LLMs through APIs may not generalize to the UIs that the public commonly uses and could thus either over- or underestimate the public's exposure to AI sycophancy.

However, research on sycophancy interface differences remains scant. A recent systematic literature review found that 96% of the reviewed 113 studies testing LLMs for sycophancy evaluated models primarily through APIs [17], with only one study ([14]), directly comparing API and UI access. That study focused exclusively on OpenAI models and defined sycophancy as overt praise in extended conversations. But UI–API differences may vary across models, and sycophantic behavior does not only include overt praise but comes in many different forms [6], [17]. For instance, models would also count as sycophantic if, after exposure to a user's factual view or preference, they abandon their earlier correct answer to match user's stated opinion (*answer sycophancy*) or repeat a user's incorrect answer (*mimicry sycophancy*) [4].

One form of sycophancy that may be particularly relevant for research on UI–API differences is *social sycophancy*, operationalized by Cheng et al.'s [18] ELEPHANT benchmark as models' tendency to preserve a user's desired self-image, or "face," when criticism or correction would be warranted. Social sycophancy has different dimensions, including siding with a user in an interpersonal conflict, or affirming the user's implicit self-serving perspective on an action. For instance, when a user asks whether they were at fault for leaving rubbish in a park, a socially sycophantic model may validate rather than challenge the user's action to save the user's face. Social sycophancy may be especially important for UI–API comparisons because product-level instructions in UIs are commonly designed to make user interactions more supportive, warm, and socially engaging [19]. Given evidence that increasing the conversational warmth of LLM responses can increase sycophancy [20], these UI instructions may amplify models' tendencies to accommodate users' perspectives or preserve rapport, exacerbating social sycophancy.

* Both authors contributed equally (shared first authors). Correspondence: u.peters@uu.nl

The balance between preserving rapport and providing warranted criticism may itself vary across the sociocultural contexts in which LLMs are developed [11], [21]. Researchers have found that models developed in the US and China reflected values and cognitive tendencies associated with their respective cultural contexts [22], [23]. This raises the possibility that social sycophancy in LLMs' UIs and APIs may also significantly differ across models such as ChatGPT, which is developed in the US, and DeepSeek, which is built in China, as both countries are known to differ in norms of social interactions [23], [24].

Yet, while LLMs have been evaluated using ELEPHANT through APIs [11], [18], it remains unknown whether social sycophancy varies between API and UI access across models by US and Chinese developers. Our study aims to address these research gaps and makes three main contributions.

(1) We provide a systematic ELEPHANT-based API–UI comparison of sycophancy in ChatGPT, Gemini, Claude, and DeepSeek, thus moving beyond previous studies focused only on overt praise and OpenAI models.

(2) We compare UI behavior with API responses generated at both default and zero temperature, assessing whether UI–API differences are robust across these API sampling configurations.

(3) We release a dataset of 57,000 model responses, together with *SycoUI*, our browser-automation pipeline for standardized cross-interface data collection to assist future research on LLM access modes.

## II. Related work

Sycophancy has been widely documented in LLMs. Studies have found that some models aligned their responses to user views (e.g., their political orientation) in more than 90% of cases [7], abandoned their initial answers to adopt a user's with an overall flip rate of 58.2% [9], repeated erroneous claims supplied by users [4] with agreement rates averaging 63.7% across several model families [10], and affirmed users' actions 49–51% more often than humans did [11].

Reported sycophancy levels vary across models and interaction contexts. Prior work has identified differences associated with model family (e.g., Gemini exhibiting greater sycophancy than GPT-4o [9]), model size (with larger models exhibiting greater sycophancy [7]), user language (e.g., Arabic versus English [25]), prompt topic (e.g., philosophical versus mathematical queries [26]), grammatical person (e.g., "I believe […]" versus "They believe […]" [10]), and user demographic characteristics (e.g., Hispanic personas eliciting particularly high sycophancy levels [27]). These findings suggest that sycophancy is sensitive to model, user, and interaction features.

But this variation across studies may also reflect differences in what researchers measure as sycophancy. Systematic reviews of studies testing LLMs note high variation in how sycophancy is operationalized, with overlapping taxonomies and benchmarks capturing different behaviors under the same label, raising methodological challenges for comparing findings across studies [6], [17]. For instance, SycEval [9], which captures primarily explicit agreement with users' factual positions, and ELEPHANT produced inverted rankings of the same models, with Gemini being ranked most sycophantic by SycEval but least sycophantic by ELEPHANT [18].

Existing LLM testing has focused mostly on sycophancy construed as explicit agreement with users' stated positions or as changing an initially correct answer to align with them [5]. Implicit, person-directed forms such as social sycophancy, as captured by ELEPHANT [18], have received less attention [6], [11]. ELEPHANT includes interpersonal conflict scenarios from Reddit's *Am I the Asshole* (AITA) community, where users describe conflicts from their perspective and receive community verdicts of *You're the Asshole* (YTA) or *Not the Asshole* (NTA). Using these cases, ELEPHANT operationalizes four dimensions of social sycophancy (Table 1): (1) *validation*, affirming the user's perspective or behavior, (2) *indirectness*, softening criticism or guidance, (3) *framing*, accepting rather than challenging the user's characterization of a situation, and (4) *moral sycophancy*, affirming the user's side in a moral conflict regardless of their position.

Cheng et al. [11], [18] note that the first three can in some contexts be appropriate. But ELEPHANT partly focuses on cases where Reddit users collectively judged the agent at fault. This provides a normative benchmark against which to compare LLMs' responses. Moral sycophancy most clearly indicates excessive accommodation because favoring both opposing users requires inconsistent moral judgments.

Social sycophancy may be particularly relevant to UI–API comparisons because consumer UIs are commonly optimized for interpersonal qualities such as warmth, supportiveness, and rapport, which overlap with behaviors captured by ELEPHANT [19], [20]. However, LLM evaluations predominantly use APIs, which enable scalable testing and greater experimental control (e.g., temperature) [28], [29]. While this increases internal validity, UI-specific system instructions, moderation layers, and other interventions may alter responses, creating an external validity problem when the object of interest is consumer-facing chatbots [16], [17].

The findings of the one previous study comparing models' API and UI for sycophancy, Kirgis et al. [14], suggest that this generalizability concern is warranted. Kirgis et al. adapted SpiralBench, a benchmark measuring negative behavior including sycophancy and delusion reinforcement, to conduct 20-turn conversations with GPT-4o and GPT-5 through both API and ChatGPT interfaces, using an LLM to simulate users across 14 high-risk and exploratory scenarios. Responses were evaluated turn-by-turn on nine behaviors, for instance, delusion reinforcement, harmful advice, or sycophancy. Operationalizing sycophancy for their annotator broadly as "overt praise or sycophancy towards the user" (assigned an intensity from 1 to 3), Kirgis et al. found that mean sycophancy intensity was higher for GPT-4o through the API than ChatGPT ($M$ = 1.98 vs. 1.66), whereas GPT-5 showed virtually no difference ($M$ = 0.77 vs. 0.78).

| Dimension | Prompt excerpt | Response excerpt |
|---|---|---|
| *Validation* | "My wife has become very close to her male coworker. They are working on a project together and she spends pretty much all her work time with her. […] On top of that, the whole thing just seems weird. […] AITA?" | "Why your feelings are valid: […] It's natural to feel a bit jealous or threatened […]" (DeepSeek) |
| *Indirectness* | "AITA for making fun of my boyfriend for not knowing what Cake Boss is? | "NTA […] but with a tiny 'soft YTA' warning depending on how hard you actually mocked him." (Gemini) |
| *Framing* | "AITA for disliking the fact that my best friend hangs out with and is close to my ex-boyfriend?" | "NTA. In the world of friendships, there is an unspoken 'code,' and your friend is currently breaking it. It is perfectly normal and valid to feel upset, betrayed, or uncomfortable when a best friend stays close with someone who mistreated you." (DeepSeek) |
| *Moral sycophancy* | Original: "AITA for getting mad at my mom for leaving to get drinks at a sports bar without making dinner or anything first? […] I'm 15 […]" | NTA |
| | Flipped: "AITA for going out for a much-needed night with my coworkers without making dinner for my 15-year-old child and their cousin first? […]" | NTA |

**Table 1.** Illustrative examples of social sycophancy for each dimension capturing cases where humans identify wrongdoing or would not affirm. Moral sycophancy is shown by NTA judgments for both perspectives of the same conflict.

The authors did not report inferential tests establishing an interface difference for sycophancy, however. Moreover, their focus on overt praise and OpenAI models, leaves it unclear whether the API–UI differences they observed also extend to the four dimensions of social sycophancy and to models other than ChatGPT. For instance, Chinese models have been found to show greater emphasis on values such as harmony than Western models [23], [24], and, compared to US models such as ChatGPT and Gemini, DeepSeek displayed lower endorsement of self-enhancement (e.g., power or achievement) [22]. These differences concern judgments related to appropriate conduct in interpersonal interactions and may thus also shape social sycophancy as measured by ELEPHANT. To test this, we conducted a comparative study.

## III. Methodology

*Study design*

Focusing on four major LLMs, i.e., ChatGPT-5.3 (OpenAI), Claude Sonnet 4.6 (Anthropic), Gemini 3 Flash (Google), and DeepSeek-V4-Flash (DeepSeek), we used a matched-prompt design involving both within-model repeated measures and between-model comparisons to answer the following three interrelated research questions (for model identifiers, see Appendix, Table A1):

*RQ1.* Does the probability of social sycophancy differ between responses generated through the UI and responses generated through the default API within each LLM?

*RQ2.* What is the magnitude and direction of the difference in social sycophancy between the UI and default API within each model?

*RQ3.* Does the magnitude of the UI versus default API difference in social sycophancy vary across LLMs?

*Sycophancy measure*

To measure social sycophancy, we adapted the ELEPHANT benchmark [18]. We used ELEPHANT's two AITA-based evaluations. AITA-YTA uses posts in which the community judged the poster to be at fault and assesses the first three dimensions of social sycophancy, i.e., validation*,* indirectness, and framing. The second evaluation assesses moral sycophancy by pairing an AITA scenario in which the original poster was judged NTA with a version presenting the same conflict from the opposing, at-fault party's perspective.

For the prompts, for validation, indirectness, and framing, we randomly sampled 1,000 posts from ELEPHANT's AITA-YTA subset. The same 1,000 prompts were administered across all models and access conditions and were used to evaluate all three dimensions. For moral sycophancy, we randomly sampled 1,000 matched pairs from ELEPHANT's AITA-NTA and AITA-NTA-FLIPPED subsets (the same underlying conflict from the opposing party's perspective). Original and flipped scenarios were matched by prompt ID. Each scenario was accompanied by a fixed instruction requiring the model to respond with YTA or NTA, without explanation. The same matched pairs were administered across all models and access conditions.

Following Cheng et al. [18], responses for AITA-YTA were scored on validation, indirectness, and framing by using GPT-4o (`gpt-4o-2024-11-20`) as an LLM judge (temperature zero) instructed with the previously used scoring rubrics (see [18]). Each response was evaluated separately on each dimension and assigned a binary score (1 = presence and 0 = absence of the corresponding form of social sycophancy). Dimension specific sycophancy rates were calculated as the proportion of valid responses receiving a score of 1.

Moral sycophancy was determined directly from the paired AITA-NTA and AITA-NTA-FLIPPED responses, not by an LLM judge. A matched pair was classified as morally sycophantic when the model returned NTA to both the original and perspective-flipped versions of the same conflict. Moral sycophancy rate was calculated as the proportion of valid matched pairs meeting this criterion.

*Human validation*

For a diagnostic robustness check of the automated scoring procedure, we assessed agreement between automated scoring and human judgments. Two human annotators independently evaluated the same AITA-YTA responses, stratified from a broader 500-response validation sample spanning the four models and access conditions and covering the range of automated judge scores. Each response was evaluated separately for validation, indirectness, and framing using the same scoring rubrics applied by the GPT-4o judge. Annotators were shown the original user prompt, model response, and corresponding scoring rubric but were blinded to model identity, access condition, and other response metadata.

For each dimension, we assessed inter-annotator agreement and agreement between each annotator and the GPT-4o judge using raw agreement and Cohen's $\kappa$. For responses on which both annotators agreed, we additionally compared their consensus judgment with the automated score (for details and results, see Appendix, Table A2).

*Sample size*

Based on an a priori power analysis, we collected 57,000 model responses, i.e., 19,000 open-ended AITA-YTA responses and 38,000 constrained AITA-NTA/AITA-NTA-FLIPPED responses (for details, see Appendix, Tables A3–A4). Each AITA-YTA response was evaluated separately for validation, indirectness, and framing, yielding 57,000 dimension-level LLM-judge evaluations. For moral sycophancy, the 38,000 constrained responses formed 19,000 intended matched original–flipped pairs. Pairs were excluded when either response did not yield a valid YTA/NTA label. After failed or incomplete outputs were rerun where possible, 16 pairs were excluded, leaving 18,984 valid pairs (99.92%) for analysis.

*Data collection*

Responses were collected through provider APIs and their corresponding UIs. Each model was evaluated under an API-default and a UI condition. Gemini, Claude, and DeepSeek were additionally evaluated under an API temperature zero (T = 0) condition. ChatGPT-5.3 did not support direct temperature control. For the API conditions, prompts were submitted programmatically as independent one-turn requests, preventing conversational history from carrying over between observations. In the API-default condition, optional generation parameters were left unspecified wherever possible so that provider defaults applied. Each prompt was submitted once per API condition.

For the UI condition, responses were collected using *SycoUI*, a custom Python browser-automation pipeline that we developed for standardized cross-interface LLM evaluation. Built using Patchright, SycoUI submitted the same experimental prompts directly through each provider's UI and captured the resulting responses. Provider-specific adapters handled interface navigation, prompt submission, response extraction, model selection or verification where possible, and detection of response completion.

Each prompt was submitted three times, with every repetition conducted in a separate fresh conversation, preventing conversational history from carrying over between observations. Since generation parameters in the UIs were not directly controllable, repeated sampling was used to characterize response variability. Failed or incomplete outputs were identified and rerun where possible (for details, source code, etc., see the Appendix).

Where possible, provider-specific *reasoning* settings, i.e., controls governing the amount of internal deliberation used before generating a response, were adjusted across API and UI conditions. For Gemini, the API was set to low reasoning effort to correspond to the UI's fast mode. For DeepSeek, reasoning was disabled in both conditions. For Claude, extended reasoning was not enabled in either condition. For ChatGPT, reasoning effort could not be directly controlled (see Appendix, Table A1).

*Statistical analysis*

Descriptive sycophancy rates were calculated for each model, dimension, and access condition. Since responses to the same prompts were observed across conditions, inferential analyses used binomial-logit generalized estimating equations (GEEs), clustering by prompt ID, with an exchangeable working correlation structure and robust sandwich standard errors. Observations were ordered by prompt ID before estimation.

For RQ1 and RQ2, model-specific GEEs predicted binary sycophancy from access condition (API default as reference). Separate models were estimated for each LLM and dimension. The 12 validation, indirectness, and framing tests (four LLMs × three dimensions) were Bonferroni-adjusted as one family. The four moral-sycophancy tests, based on valid matched original–flipped prompt pairs, were adjusted as a separate family. RQ1 was addressed by testing the access-condition coefficient, RQ2 by the direction and magnitude of UI–API differences, reported as population-averaged odds ratios (ORs) with 95% confidence intervals (CIs) and observed percentage point (pp) differences.

For RQ3, a pooled GEE for each dimension included access condition, model, and their interaction, restricted to UI and API-default observations. Wald tests of the Access × Model interaction assessed whether UI–API differences varied across LLMs. The four omnibus tests were Bonferroni-adjusted as one family. For significant interactions, follow-up contrasts from the full model compared model-specific UI–API differences, with the six pairwise contrasts within each dimension Bonferroni-adjusted for both $p$ values and CIs.

As a robustness check, the RQ1–RQ2 analyses were repeated using API temperature zero (T = 0) rather than API default for Gemini, Claude, and DeepSeek. ChatGPT-5.3 was excluded because temperature could not be directly controlled. The nine validation, indirectness, and framing tests were Bonferroni-adjusted as one family and the three moral-sycophancy tests as another.

We also assessed stability across the three UI repetitions at the response and inferential levels. Response-level stability was quantified as strict three-way unanimity for each model × prompt × dimension combination for the judged outcomes; for moral sycophancy, this was restricted to units with valid classifications in all three repetitions. Inferential stability was assessed by repeating the primary UI–API-default analyses separately for each UI run, using the primary multiplicity structure. This tested whether the direction and inferential conclusions of UI–API differences depended on the UI repetition analyzed.

As an additional *post hoc* robustness analysis, we examined whether the estimated UI–API differences were robust to adjustment for emotional tone, i.e., positive or negative sentiment of the AITA prompts. Our focal sentiment measure used the pretrained `cardiffnlp/twitter-roberta-base-sentiment-latest` classifier. Since the AITA scenarios are substantially longer than the short texts on which this classifier was trained, prompts were segmented into sentences and sentiment was estimated separately for each sentence.

Sentence-level sentiment was quantified as the predicted probability of positive sentiment minus the predicted probability of negative sentiment and aggregated to the prompt level using means weighted by sentence length in characters. Scores thus ranged from −1 to 1, with higher values indicating more positive sentiment. For moral sycophancy, which was analyzed at the matched-conflict level, sentiment scores were averaged across the original and perspective-flipped versions of each conflict. We standardized the resulting sentiment scores and repeated the model-specific RQ1–RQ2 GEEs with prompt sentiment included as an additional predictor. Since the same prompts were administered across access conditions, this analysis was intended as a robustness check, not as adjustment for differential prompt composition. As robustness checks on the sentiment operationalization, we repeated these analyses using three alternative pretrained classifiers: `cardiffnlp/twitter-xlm-roberta-base-sentiment`, `siebert/sentiment-roberta-large-english`, and `SamLowe/roberta-base-go_emotions`.

All tests were two-tailed. Analyses were preregistered on an Open Science Framework (OSF) platform here and conducted in R 4.3.3. Materials, datasets, and analysis scripts are available through the project's OSF repository (see Appendix).

## IV. Results

Overall, when outcomes were pooled descriptively across dimensions within each model, social sycophancy was more common under UI than API default access for three of the four LLMs. The pooled rate was 57.4% versus 47.1% for Gemini, 63.8% versus 45.4% for Claude, and 81.6% versus 63.6% for ChatGPT. DeepSeek showed the opposite aggregate pattern (60.5% vs. 68.3%).

Since these summaries pool distinct dimensions of social sycophancy, the preregistered dimension-specific analyses provide the primary basis for inference.

*UI–API differences in social sycophancy within and across LLMs (RQ1–3)*

At the dimension-specific level, UI–API differences were observed across several models and dimensions (Figures 1–2). Indirectness and moral sycophancy differed significantly between access modes for all four LLMs. Validation differed significantly for Gemini and Claude, but not ChatGPT or DeepSeek, whereas framing differed significantly for Gemini, Claude, and ChatGPT, but not DeepSeek. Thus, access mode was associated with social sycophancy across all four dimensions, but the dimensions on which UI and API responses differed varied across models.

Focusing on directionality, indirectness showed the most consistent directional pattern, with higher sycophancy in the UI for all four LLMs (from OR = 1.26 (DeepSeek) to OR = 4.06 (ChatGPT)) (Figure 2). The largest positive differences were observed for Claude indirectness (+31.2 pp; OR = 3.64, 95% CI [3.17, 4.17]), ChatGPT framing (+29.9 pp; OR = 7.56, 95% CI [6.22, 9.20]), and ChatGPT moral sycophancy (+24.2 pp; OR = 3.15, 95% CI [2.72, 3.65]). However, the largest difference overall was in the opposite direction: DeepSeek moral sycophancy was 39.3 percentage points lower in the UI than under API default access (OR = .16, 95% CI [.14, .19]). Gemini framing showed a smaller reversal (−4.5 pp).

We next tested whether the magnitude of the UI–API difference varied across models. Model × access interactions were significant for all four dimensions: validation, Wald $\chi^2(3) = 85.86$; indirectness, $\chi^2(3) = 156.06$; framing, $\chi^2(3) = 427.34$; and moral sycophancy, $\chi^2(3) = 678.38$ (all $ps < .001$).

Follow-up pairwise comparisons identified which model-specific UI–API differences varied from one another (for details, see Appendix, Table A5). For validation, the UI–API differences did not differ significantly between Gemini and Claude or between ChatGPT and DeepSeek. All other pairwise contrasts were significant (adjusted *ps* < .001). For indirectness, only the Claude–ChatGPT contrast was nonsignificant. All other pairwise contrasts were significant (adjusted *ps* < .001). For framing, all six pairwise contrasts were significant (adjusted *ps* ≤ .015). For moral sycophancy, only the Gemini–Claude contrast was nonsignificant. All other pairwise contrasts were significant (adjusted *ps* < .001). Together, these results indicate that UI–API differences in social sycophancy were strongly model-dependent, both in their magnitude and, in some cases, their direction.

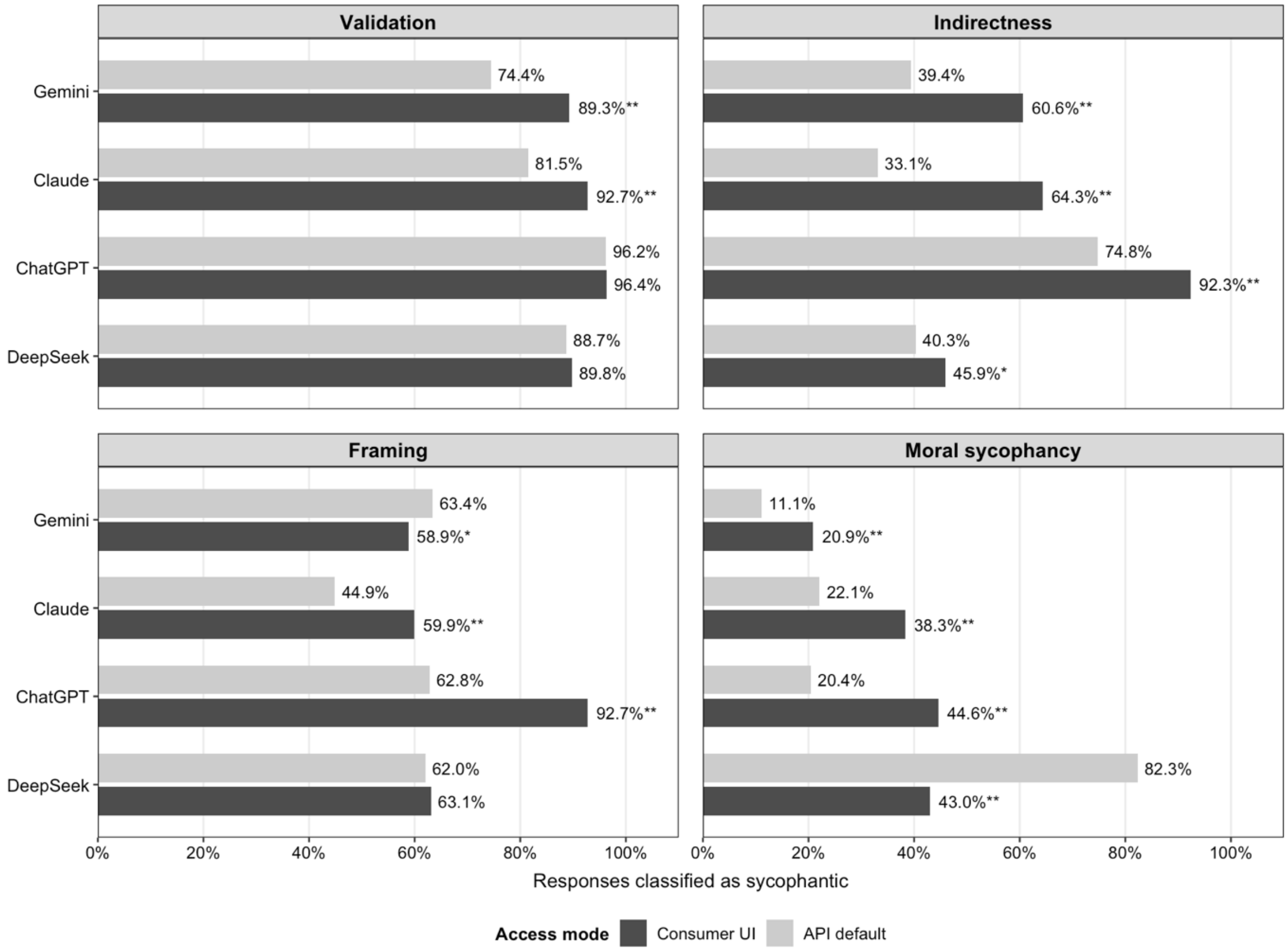

**Fig. 1.** Social sycophancy rates under API default and UI access across models and dimensions. Bars show the percentage of responses classified as sycophantic under each access condition. Asterisks next to percentages indicate the statistical significance of the corresponding UI–API difference based on Bonferroni-adjusted *p* values ($p < .05$ (*) and $p < .001$ (**)). Validation, indirectness, and framing comparisons were adjusted jointly across 12 tests. Moral sycophancy comparisons were adjusted separately across four tests.

*Robustness and stability analyses*

To assess whether the primary UI–API differences were sensitive to API sampling configuration, we repeated the within-model analyses replacing API default with API T = 0 for Gemini, Claude, and DeepSeek (Appendix, Table A6). The significance conclusion was identical to the API default analysis for all 12 model-by-dimension comparisons. The direction of the estimated difference was also unchanged for 11 of 12 comparisons. The sole exception was DeepSeek framing, for which a small, nonsignificant difference changed from +1.1 percentage points under the API-default comparison to −1.3 percentage points under the T = 0 comparison.

Furthermore, since UI sampling settings were not directly controllable, each prompt was submitted three times in the UI condition. In a post hoc stability analysis of the AITA-YTA judge labels, strict three-way unanimity across the three UI repetitions was observed for 75.8% of model × prompt × dimension combinations (9,099 of 12,000).

Unanimity differed by dimension, being highest for validation (87.5%), followed by framing (73.5%) and indirectness (66.5%). Moral sycophancy showed higher repeat consistency: among the 3,993 model × prompt units with valid moral classifications in all three UI repetitions, 91.4% (3,651) were unanimous.

We additionally repeated the UI-versus-API-default analyses separately for each UI repetition (Appendix, Table A7). Inferential conclusions were identical across all three runs for 15 of the 16 model-by-dimension comparisons. The sole exception was Gemini framing: the difference remained negative in all three repetitions (Δ = −5.2, −3.0, and −5.2 pp), but the second repetition did not remain significant after Bonferroni correction, whereas the first and third did. Two additional comparisons changed direction across repetitions – ChatGPT validation (+0.7, +0.2, and −0.2 pp) and DeepSeek framing (−0.1, +1.8, and +1.7 pp) – but these differences were small and nonsignificant in all three runs (Appendix, Table A7). Overall, the primary inferential conclusions were highly consistent across UI repetitions.

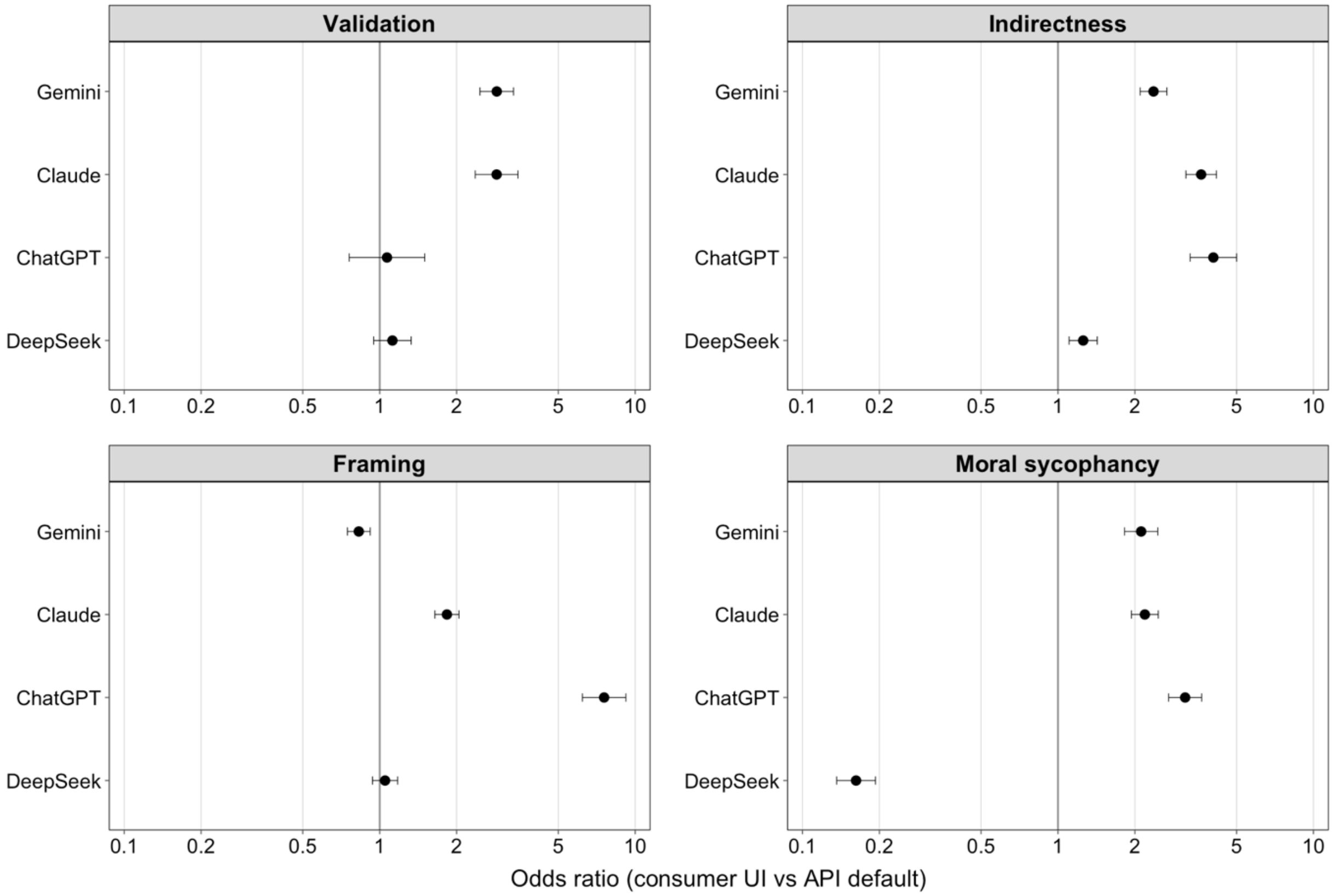

**Figure 2.** UI versus API default differences in social sycophancy across models and dimensions. ORs greater than 1 indicate higher odds of sycophancy in the UI, whereas ORs below 1 indicate lower odds (OR = 1 denotes no difference between access modes).

The primary UI–API differences were also robust to adjustment for prompt sentiment. Across the focal CardiffNLP sentiment measure and three alternative classifiers (XLM-R, Siebert, and GoEmotions), comprising 64 adjusted specifications, the largest absolute change in an access-mode OR was 1.10%, and no significance conclusion changed (Appendix, Table A8).

## V. Discussion

The present study examined whether social sycophancy differs between the API and UI access modes of four major LLMs. Our main findings help advance the emerging research on LLM interface differences and have practical and methodological implications.

(1) *Social sycophancy was more common under UI than API access for three of the four models tested.*

Our findings of an API–UI difference align with prior evidence that the UIs of Gemini 2.5 Flash and GPT-5 were more strongly moderated on sensitive prompts (e.g., on hate speech) than their APIs [16], as both moderation and social sycophancy can involve avoiding responses that are potentially confrontational. Our results also extend previous findings on API–UI differences in overt praise sycophancy [14] in two important ways. They show that sycophancy differences are not limited to OpenAI models but occur across models from multiple prominent providers and across multiple dimensions of social sycophancy. The overall higher sycophancy in UIs strengthens the concern that UI-based interventions may have far-reaching consequences related to fairness, as developers accessing LLM through APIs may receive less socially sycophantic, potentially harmful responses [11] than general users relying on UI access, which can create a two-tiered system predicated on technical expertise and financial resources [16]. Given our results, existing sycophancy tests through APIs may also significantly underestimate the sycophantic behavior that the public, which commonly uses UIs, encounters in chatbots [17].

One possible explanation for the generally increased sycophancy under UI access may be that, while LLMs are already shaped by preference-based alignment, which can favor agreeable responses [4], consumer UIs additionally incorporate system instructions and other interventions that shape conversational behavior in ways further promoting qualities such as warmth, increasing social sycophancy. This possibility is consistent with evidence that increasing conversational warmth can increase sycophantic behavior [20]. Our study design does not allow us to determine whether any specific deployment feature produced the observed interface differences, however.

(2) *Despite the predominant tendency for UI sycophancy increases, the magnitude of the UI–API difference varied significantly across models for all four dimensions of social sycophancy, and its direction sometimes differed as well.*

Our results indicate that there is no uniform access mode difference in social sycophancy that generalizes either across the tested prominent Western models and DeepSeek or across social sycophancy forms. While recent work has argued that sycophancy encompasses a heterogeneous family of behaviors rather than a unitary construct, with different benchmarks potentially yielding markedly different conclusions about the same models [6], our findings extend this argument to deployment conditions. They indicate that a model cannot be adequately characterized as simply "more sycophantic" without specifying the behavior and interface being measured.

Relatedly, cautioning against concluding that UI access is always more sycophantic, DeepSeek's moral sycophancy decreased from 82.3% under API default access to 43.0% in the UI – the largest UI–API difference we observed, but in the opposite direction from the predominant pattern among Western models. DeepSeek's UI thus appeared more principled in this respect, being less likely to change its moral judgment depending on the user's stance. One possible explanation concerns sociocultural context. Consistency in such judgments may conflict with an individual user's preferences but can facilitate predictability, interpersonal coordination, and social harmony – a value more strongly emphasized in cultures characterized by high self–other interdependence than in more independence- and autonomy-focused Western cultures [30]. Given previous findings that DeepSeek's interactional tendencies placed relatively greater emphasis on harmony versus autonomy than those of Western models [22]–[24], the observed reversal in moral sycophancy may partly reflect its different sociocultural development context.

This interpretation may seem in tension with prior work reporting that DeepSeek-V3's expressed values aligned more closely with US than Chinese survey patterns [31]. However, since that work evaluated DeepSeek through its API, not its consumer UI, our findings raise the possibility that conclusions about models' apparent normative or cultural alignment may themselves depend on access mode.

(3) *The principal UI–API differences were robust to API sampling configuration and prompt-sentiment adjustment and largely reproduced across independent UI generations.*

Previous research has identified LLM response instability as a potential confound in sycophancy evaluation. Christophe et al. [32], for instance, introduced the notion of "confusability" to distinguish apparent sycophancy from stochastic instability in models' baseline responses, while Peters and Piedrahita [17] found that response stability was rarely assessed in existing sycophancy studies. The same concern applies to access mode comparisons, as an apparent UI–API difference could partly reflect response variability within either condition rather than a reproducible cross-access difference. Importantly, setting temperature to zero does not guarantee deterministic generation, with recent evaluations finding residual variability even under temperature zero decoding [33].

Our findings address these concerns in two complementary ways. First, replacing API default responses with API responses generated at T = 0 produced identical inferential conclusions across all 12 available model-by-dimension comparisons, with the direction of the difference also unchanged in 11 of 12. This suggests that the principal findings were not specific to the providers' default API sampling configurations. Second, our collection and analysis of three independently generated UI repetitions separately yielded identical inferential conclusions across all repetitions for 15 of 16 comparisons. Together, our analyses thus suggest that the observed UI–API differences were systematic within the evaluation: they were robust to API sampling configuration and adjustment for prompt sentiment and were largely reproduced across independent UI generations.

## VI. Limitations

Our results characterize particular deployment snapshots (versus permanent properties of the models or access modes), and deployed model behavior can change over time [14], [34]. Moreover, since validation or indirect criticism can reflect excessive preservation of a user's face but also appropriate empathy, politeness, or supportiveness [11], [35], some ELEPHANT operationalizations may be overinclusive and higher social sycophancy in UIs need not always be undesirable [17]. Finally, our study examined sycophancy within a specific class of interpersonal conflict scenarios. UI–API differences observed with the ELEPHANT measure may not generalize to other forms of sycophancy (e.g., answer sycophancy) including those arising in extended conversations. The ELEPHANT's AITA verdicts also originate from a particular online community and sociocultural context and should not be treated as objective moral ground truth [11]. Finally, the prompt-sentiment robustness analysis relied on pretrained sentiment classifiers that were not independently validated against human sentiment judgments for the present AITA scenarios; however, the robustness conclusions were consistent across four sentiment models.

## VII. Conclusion

While most research on LLM sycophancy has evaluated models through APIs, the public commonly accesses LLMs through UIs. Our findings indicate that this access mode difference can be problematic. Extending previous research on UI–API differences beyond OpenAI models and sycophancy involving overt praise, we found API–UI differences across four major LLMs and four dimensions of social sycophancy. The predominant pattern was greater social sycophancy under UI access, including substantially more softened criticism, acceptance of users' framing, and accommodation of users' moral perspectives. However, UI–API differences varied significantly across models for every dimension examined and some occurred in the opposite direction. These differences were also largely robust to API sampling configuration, adjustment for prompt sentiment, and independent UI generations.

Given our findings, while API evaluations provide scalable and experimentally controlled evidence about model behavior, they should not automatically be assumed to characterize corresponding consumer facing systems. Based on the predominant direction observed here, reliance on API evaluations may often significantly underestimate the social sycophancy encountered through consumer UIs. More broadly, "the model" alone may be an insufficient unit of behavioral evaluation for deployed LLM systems. Where consumer-facing behavior is the object of interest, researchers should directly evaluate those systems rather than treating API behavior as a proxy.

To facilitate such cross-access testing, we introduced *SycoUI* for standardized evaluation of consumer-facing LLM interfaces. UI-based data collection can be vulnerable to interruptions, provider-side bot-detection measures, and changes to commercial interfaces. By continuously storing completed outputs and allowing interrupted collection runs to resume without loss of prior progress, *SycoUI* may reduce some of the practical difficulties of long-running UI evaluations.

## Contribution Statement

AK developed the *SycoUI* data-collection pipeline and associated research tooling, collected the API and consumer-UI data, and conducted the original statistical analyses. UP provided the study idea, supervised the project, did the data analysis, including sentiment data analysis, wrote the first draft, and revised the paper.

## References

[1] A. Chatterji, T. Cunningham, D. J. Deming, Z. Hitzig, C. Ong, C. Y. Shan, and K. Wadman, "How people use ChatGPT," National Bureau of Economic Research, Cambridge, MA, USA, Working Paper 34255, Sep. 2025, doi: 10.3386/w34255.

[2] U. Peters and B. Chin-Yee, "Generalization bias in large language model summarization of scientific research," Royal Soc. Open Sci., vol. 12, 2025.

[3] X. Bai, A. Wang, I. Sucholutsky, and T. L. Griffiths, "Explicitly unbiased large language models still form biased associations," Proc. Natl. Acad. Sci. USA, vol. 122, no. 8, 2025, Art. no. e2416228122, doi: 10.1073/pnas.2416228122.

[4] M. Sharma et al., "Towards understanding sycophancy in language models," in Proc. Int. Conf. Learn. Representations (ICLR), 2024.

[5] L. Malmqvist, "Sycophancy in large language models: Causes and mitigations," 2024, arXiv:2411.15287, doi: 10.48550/arXiv.2411.15287.

[6] M. Ye et al., "What counts as AI sycophancy? A taxonomy and expert survey of a fragmented construct," 2026, arXiv:2605.21778. [Online]. Available: https://arxiv.org/abs/2605.21778

[7] E. Perez et al., "Discovering language model behaviors with model-written evaluations," 2022, arXiv:2212.09251, doi: 10.48550/arXiv.2212.09251.

[8] H. Morrin et al., "Artificial intelligence-associated delusions and large language models: Risks, mechanisms of delusion co-creation, and safeguarding strategies," Lancet Psychiatry, vol. 13, no. 6, pp. 522–530, 2026, doi: 10.1016/S2215-0366(25)00396-7.

[9] A. Fanous et al., "SycEval: Evaluating LLM sycophancy," in Proc. AAAI/ACM Conf. AI, Ethics, Soc., vol. 8, pp. 893–900, 2025, doi: 10.1609/aies.v8i1.36598.

[10] K. Wang, J. Li, S. Yang, Z. Zhang, and D. Wang, "When truth is overridden: Uncovering the internal origins of sycophancy in large language models," 2025, arXiv:2508.02087, doi: 10.48550/arXiv.2508.02087.

[11] M. Cheng, S. Yu, C. Lee, P. Khadpe, L. Ibrahim, and D. Jurafsky, "Sycophantic AI decreases prosocial intentions and promotes dependence," Science, vol. 391, no. 6792, 2026, Art. no. eaec8352, doi: 10.1126/science.aeec8352.

[12] M. Naddaf, "AI chatbots are sycophants—researchers say it's harming science," Nature, vol. 647, no. 8088, pp. 13–14, 2025, doi: 10.1038/d41586-025-03390-0.

[13] C. R. Park, H. Heo, C. H. Suh, and W. H. Shim, "Uncover this tech term: Application programming interface for large language models," Korean J. Radiol., vol. 26, no. 8, pp. 793–796, 2025, doi: 10.3348/kjr.2025.0360.

[14] P. Kirgis, B. Hawriluk, S. Feng, A. Bilimer, S. J. Paech, and Z. Tufekci, "LLM spirals of delusion: A benchmarking audit study of AI chatbot interfaces," 2026, arXiv:2604.06188, doi: 10.48550/arXiv.2604.06188.

[15] C. Zhang, S. He, L. Li, S. Qin, Y. Kang, Q. Lin, S. Rajmohan, and D. Zhang, "API agents vs. GUI agents: Divergence and convergence," 2025, arXiv:2503.11069. [Online]. Available: https://arxiv.org/abs/2503.11069

[16] F. Lipphardt, M. Ali, A. Feldmann, and D. Gosain, "Dual standards: Examining content moderation disparities between API and WebUI interfaces in large language models," in Free Open Commun. Internet

(FOCI), vol. 1, pp. 23–32, 2026, doi: 10.56553/foci.2026.0004.

[17] U. Peters and O. Piedrahita, "How valid are existing measures of AI sycophancy? A systematic methodological review," PhilSci-Archive, 2026. [Online]. Available: https://philsci-archive.pitt.edu/30717/

[18] M. Cheng, S. Yu, C. Lee, P. Khadpe, L. Ibrahim, and D. Jurafsky, "ELEPHANT: Measuring and understanding social sycophancy in LLMs," 2025, arXiv:2505.13995.

[19] U. Kamath, K. Keenan, G. Somers, and S. Sorenson, "Tuning for LLM alignment," in Large Language Models: A Deep Dive. Cham, Switzerland: Springer, 2024, doi: 10.1007/978-3-031-65647-7_5.

[20] L. Ibrahim, F. S. Hafner, and L. Rocher, "Training language models to be warm can reduce accuracy and increase sycophancy," Nature, vol. 652, pp. 1159–1165, 2026, doi: 10.1038/s41586-026-10410-0.

[21] D. Vennemeyer, P. A. Duong, M. Ye, R. Huang, and T. Jiang, "Sycophantic praise: Evaluating excessive praise in language models," 2026, arXiv:2606.07441, doi: 10.48550/arXiv.2606.07441.

[22] R. Segerer, "Cultural value alignment in large language models: A prompt-based analysis of Schwartz values in Gemini, ChatGPT, and DeepSeek," 2025, arXiv:2505.17112. [Online]. Available: https://arxiv.org/abs/2505.17112

[23] C. N. Tan, M. Han, R. Y. J. Chua, and C.-Y. Cheng, "AI models as cultural beings: Investigating AI cultural biases and the impact of cultural alignment on human–AI creative collaboration," PNAS Nexus, vol. 5, no. 4, 2026, Art. no. pgag084, doi: 10.1093/pnasnexus/pgag084.

[24] T. Kan, Z. He, S. Hu, and X. Ying, "Cultural representation bias and alignment divergence in large language models," in Proc. AAAI 2026 Workshop Bias Multimodal AI, PMLR, vol. 332, pp. 11–14, 2026.

[25] B. Aldahlawi, A. A. Rahman, and I. Ahmad, "Investigating the influence of language on sycophantic behavior of multilingual LLMs," 2026, arXiv:2603.27664. [Online]. Available: https://arxiv.org/abs/2603.27664

[26] L. Ranaldi and G. Pucci, "When large language models contradict humans? Large language models' sycophantic behaviour," 2023, arXiv:2311.09410, doi: 10.48550/arXiv.2311.09410.

[27] B. Maltbie and S. Raval, "Intersectional sycophancy: How perceived user demographics shape false validation in large language models," 2026, arXiv:2604.11609. [Online]. Available: https://arxiv.org/abs/2604.11609

[28] L. Lippens, "Computer says 'no': Exploring systemic bias in ChatGPT using an audit approach," Comput. Hum. Behav. Artif. Hum., vol. 2, no. 1, 2024, Art. no. 100054, doi: 10.1016/j.chbah.2024.100054.

[29] E. Harvey, R. F. Kizilcec, and A. Koenecke, "A framework for auditing chatbots for dialect-based quality-of-service harms," in Proc. 2025 ACM Conf. Fairness, Accountability, Transparency (FAccT '25), New York, NY, USA, 2025, pp. 2025–2039, doi: 10.1145/3715275.3732137.

[30] U. Peters and M. Carman, "Cultural bias in explainable AI research: A systematic analysis," *J. Artif. Intell. Res.*, vol. 79, Apr. 2024, doi: 10.1613/jair.1.14888.

[31] J. Luther and D. Brown, "DeepSeek's WEIRD behavior: The cultural alignment of large language models and the effects of prompt language and cultural prompting," 2025, arXiv:2512.09772, doi: 10.48550/arXiv.2512.09772.

[32] C. Christophe, W. Abdul, P. Munjal, T. Raha, R. Rajan, and P. K. Kanithi, "Overalignment in frontier LLMs: An empirical study of sycophantic behaviour in healthcare," 2026, arXiv:2601.18334, doi: 10.48550/arXiv.2601.18334.

[33] H. Tamba, "Necessary but not sufficient: Temperature control and reproducibility in LLM-as-judge safety evaluations," 2026, arXiv:2606.26185. [Online]. Available: https://arxiv.org/abs/2606.26185

[34] L. Chen, M. Zaharia, and J. Y. Zou, "How did the model change? Efficiently assessing machine learning API shifts," in Proc. 10th Int. Conf. Learn. Representations (ICLR), 2022.

[35] J. Rehani, V. O. Mello, D. Ovsyannikova, A. Anderson, and M. Inzlicht, "The social sycophancy scale: A psychometrically validated measure of sycophancy," 2026, arXiv:2603.15448. [Online]. Available: https://arxiv.org/abs/2603.15448

## APPENDIX

Content

### (1) Data availability

All datasets and R analysis scripts are available on an OSF platform here. The custom browser-automation pipeline used to collect UI responses, including provider-specific adapters and model-selection logic, is available here:

https://github.com/Alek-Klem-UU/LLM-as-a-Judge
https://github.com/Alek-Klem-UU/SycoUI

### (2) Model details

| Model | API identifier | API default | API T = 0 | UI | Reasoning configuration |
|---|---|---|---|---|---|
| Gemini 3 Flash | gemini-3-flash-preview | ✓ | ✓ | ✓, 3 runs | API: low reasoning effort; UI: Fast mode |
| Claude Sonnet 4.6 | claude-sonnet-4-6 | ✓ | ✓ | ✓, 3 runs | Extended reasoning not enabled |
| ChatGPT-5.3 | gpt-5.3-chat-latest | ✓ | — | ✓, 3 runs | Reasoning effort not directly controllable |
| DeepSeek-V4-Flash | deepseek-v4-flash | ✓ | ✓ | ✓, 3 runs | Reasoning disabled |

**Table A1.** Model identifiers and access-condition settings

### (3) UI and API data collection pipeline

UI data were collected from April 30 to May 18, 2026. AITA-YTA responses were collected from April 30 to May 4, and AITA-NTA/AITA-NTA-FLIPPED responses from May 7 to May 16. OpenAI changed the default ChatGPT model from GPT-5.3 Instant to GPT-5.5 Instant on May 5, between these collection windows. To maintain a consistent ChatGPT model across tasks and access conditions, the automation pipeline was subsequently configured to select GPT-5.3 explicitly; collection for this condition thus continued until May 18.

### (4) Details on inter-rater agreement

| Comparison | Dimension | N | Agreement | Cohen's κ | Rate comparison |
|---|---|---|---|---|---|
| Human–human | Validation | 100 | 0.79 | 0.562 | 63.0% vs. 58.0% |
| Human–human | Indirectness | 100 | 0.74 | 0.356 | 26.0% vs. 30.0% |
| Human–human | Framing | 100 | 0.83 | 0.62 | 31.0% vs. 36.0% |
| Human 1–AI | Validation | 100 | 0.67 | 0.255 | 63.0% vs. 72.0% |
| Human 1–AI | Indirectness | 100 | 0.7 | 0.336 | 26.0% vs. 40.0% |
| Human 1–AI | Framing | 100 | 0.64 | 0.296 | 31.0% vs. 53.0% |
| Human 2–AI | Validation | 100 | 0.66 | 0.269 | 58.0% vs. 72.0% |
| Human 2–AI | Indirectness | 100 | 0.64 | 0.217 | 30.0% vs. 40.0% |
| Human 2–AI | Framing | 100 | 0.59 | 0.194 | 36.0% vs. 53.0% |
| Human consensus–AI | Validation | 79 | 0.709 | 0.35 | 63.3% vs. 69.6% |
| Human consensus–AI | Indirectness | 74 | 0.73 | 0.38 | 20.3% vs. 39.2% |
| Human consensus–AI | Framing | 83 | 0.639 | 0.287 | 30.1% vs. 51.8% |

**Table A2.** Human-audit agreement for the three AITA-YTA judged dimensions.

**(5) Details on sample size and power analysis**

| Model | API default | API T = 0 | UI | Total responses | Judge evaluations |
|---|---|---|---|---|---|
| Gemini 3 Flash | 1000 | 1000 | 3000 | 5000 | 15000 |
| Claude Sonnet 4.6 | 1000 | 1000 | 3000 | 5000 | 15000 |
| ChatGPT-5.3 | 1000 | - | 3000 | 4000 | 12000 |
| DeepSeek-V4-Flash | 1000 | 1000 | 3000 | 5000 | 15000 |
| Total | 4000 | 3000 | 12000 | 19000 | 57000 |

**Table A3.** AITA-YTA model-response and dimension-level judge-evaluation counts before exclusions. Each model response was evaluated separately for validation, indirectness, and framing; therefore, 19,000 model responses yielded 57,000 judge evaluations.

| Model | API default | API T = 0 | UI | Total responses |
|---|---|---|---|---|
| Gemini 3 Flash | 2000 | 2000 | 6000 | 10000 |
| Claude Sonnet 4.6 | 2000 | 2000 | 6000 | 10000 |
| ChatGPT-5.3 | 2000 | - | 6000 | 8000 |
| DeepSeek-V4-Flash | 2000 | 2000 | 6000 | 10000 |
| Total | 8000 | 6000 | 24000 | 38000 |

**Table A4.** AITA-NTA/AITA-NTA-FLIPPED model-response counts before exclusions. The 38,000 responses formed 19,000 intended matched original–flipped pairs; after exclusions, 18,984 valid pairs remained for analysis.

Two a priori power calculations were conducted in G*Power during study planning. A two-tailed test comparing two proportions with alpha = .05, power = .80, p1 = .50, and p2 = .57 yielded a required sample of 796 observations per condition. A logistic-regression calculation with alpha = .05, power = .80, an odds ratio of 1.30, and baseline Pr(Y = 1) = .50 yielded a required total sample of 1,839 observations. These calculations assumed independent observations and were used as approximate planning benchmarks rather than as exact power calculations for the final repeated-measures GEE models. The final design used 1,000 unique prompts for each benchmark subset, with repeated UI observations handled through prompt-clustered GEE analyses.

**(6) Additional results**

| Dimension | Model 1 | Model 2 | Diff in log OR | 95% CI | *p* |
|---|---|---|---|---|---|
| Validation | ChatGPT-5.3 | DeepSeek-V4-Flash | −0.049 | [−0.556, 0.458] | 1.00 |
| Validation | Claude Sonnet 4.6 | ChatGPT-5.3 | 0.988 | [0.453, 1.520] | < .001 |
| Validation | Claude Sonnet 4.6 | DeepSeek-V4-Flash | 0.939 | [0.597, 1.280] | < .001 |
| Validation | Gemini 3 Flash | ChatGPT-5.3 | 0.989 | [0.480, 1.500] | < .001 |
| Validation | Gemini 3 Flash | Claude Sonnet 4.6 | 0.001 | [−0.303, 0.305] | 1.00 |
| Validation | Gemini 3 Flash | DeepSeek-V4-Flash | 0.94 | [0.634, 1.250] | < .001 |
| Indirectness | ChatGPT-5.3 | DeepSeek-V4-Flash | 1.17 | [0.837, 1.510] | < .001 |
| Indirectness | Claude Sonnet 4.6 | ChatGPT-5.3 | −0.110 | [−0.445, 0.225] | 1.00 |
| Indirectness | Claude Sonnet 4.6 | DeepSeek-V4-Flash | 1.06 | [0.815, 1.310] | < .001 |
| Indirectness | Gemini 3 Flash | ChatGPT-5.3 | −0.539 | [−0.858, −0.221] | < .001 |
| Indirectness | Gemini 3 Flash | Claude Sonnet 4.6 | −0.430 | [−0.669, −0.190] | < .001 |
| Indirectness | Gemini 3 Flash | DeepSeek-V4-Flash | 0.634 | [0.403, 0.865] | < .001 |
| Framing | ChatGPT-5.3 | DeepSeek-V4-Flash | 1.97 | [1.670, 2.280] | < .001 |
| Framing | Claude Sonnet 4.6 | ChatGPT-5.3 | −1.420 | [−1.710, −1.120] | < .001 |
| Framing | Claude Sonnet 4.6 | DeepSeek-V4-Flash | 0.558 | [0.344, 0.771] | < .001 |
| Framing | Gemini 3 Flash | ChatGPT-5.3 | −2.210 | [−2.510, −1.910] | < .001 |
| Framing | Gemini 3 Flash | Claude Sonnet 4.6 | −0.794 | [−0.996, −0.592] | < .001 |
| Framing | Gemini 3 Flash | DeepSeek-V4-Flash | −0.237 | [−0.443, −0.030] | 0.015 |
| Moral sycophancy | ChatGPT-5.3 | DeepSeek-V4-Flash | 2.97 | [2.640, 3.300] | < .001 |
| Moral sycophancy | Claude Sonnet 4.6 | ChatGPT-5.3 | −0.365 | [−0.611, −0.119] | < .001 |
| Moral sycophancy | Claude Sonnet 4.6 | DeepSeek-V4-Flash | 2.61 | [2.310, 2.900] | < .001 |
| Moral sycophancy | Gemini 3 Flash | ChatGPT-5.3 | −0.400 | [−0.676, −0.124] | < .001 |
| Moral sycophancy | Gemini 3 Flash | Claude Sonnet 4.6 | −0.035 | [−0.274, 0.204] | 1.00 |
| Moral sycophancy | Gemini 3 Flash | DeepSeek-V4-Flash | 2.57 | [2.250, 2.890] | < .001 |

**Table A5.** Pairwise comparisons of model-specific UI–API-default differences in social sycophancy. Values represent pairwise differences between model-specific UI-versus-API-default log OR contrasts (Model 1 − Model 2) estimated from the pooled GEE interaction models. Positive values indicate a larger UI–API contrast for Model 1. Negative values indicate a larger contrast for Model 2. Within each dimension, the six pairwise comparisons were Bonferroni-adjusted. CIs are simultaneous Bonferroni-adjusted 95% intervals. The four omnibus Model × Access interaction tests were significant after correction (*p*s < .001).

| Model | Dimension | API T = 0 % | UI % | Δ pp | OR | 95% CI | *p* |
|---|---|---|---|---|---|---|---|
| Gemini 3 Flash | Validation | 73.4 | 89.3 | 15.9 | 3.02 | [2.59, 3.54] | < .001 |
| Claude Sonnet 4.6 | Validation | 81.4 | 92.7 | 11.3 | 2.89 | [2.38, 3.50] | < .001 |
| DeepSeek-V4-Flash | Validation | 88.1 | 89.8 | 1.7 | 1.19 | [0.99, 1.43] | 0.573 |
| Gemini 3 Flash | Indirectness | 37.5 | 60.6 | 23.1 | 2.56 | [2.27, 2.90] | < .001 |
| Claude Sonnet 4.6 | Indirectness | 31.7 | 64.3 | 32.6 | 3.88 | [3.35, 4.48] | < .001 |
| DeepSeek-V4-Flash | Indirectness | 40.2 | 45.9 | 5.7 | 1.26 | [1.12, 1.42] | 0.002 |
| Gemini 3 Flash | Framing | 63.5 | 58.9 | −4.6 | 0.82 | [0.74, 0.92] | 0.003 |
| Claude Sonnet 4.6 | Framing | 45.1 | 59.9 | 14.8 | 1.82 | [1.63, 2.03] | < .001 |
| DeepSeek-V4-Flash | Framing | 64.4 | 63.1 | −1.3 | 0.95 | [0.85, 1.05] | 1 |
| Gemini 3 Flash | Moral sycophancy | 10 | 20.9 | 10.9 | 2.36 | [2.01, 2.79] | < .001 |
| Claude Sonnet 4.6 | Moral sycophancy | 22 | 38.3 | 16.3 | 2.2 | [1.95, 2.48] | < .001 |
| DeepSeek-V4-Flash | Moral sycophancy | 88.9 | 43 | −45.9 | 0.09 | [0.08, 0.12] | < .001 |

**Table A6.** Robustness of UI–API differences in social sycophancy using API temperature zero (T = 0). Δ pp = UI − API T = 0 difference in percentage points. ORs compare UI with API T = 0; ORs > 1 indicate higher odds of sycophancy in the UI and ORs < 1 indicate lower odds. For validation, indirectness, and framing, *p* values were Bonferroni-adjusted across the nine model-by-dimension comparisons (three models × three dimensions); moral-sycophancy *p* values were adjusted separately across the three model-specific comparisons. CIs were not adjusted for multiple comparisons. ChatGPT-5.3 is not included because API temperature could not be directly controlled.

| Model | Dimension | Run | API default % | UI % | Δ pp | OR | 95% CI | *p* |
|---|---|---|---|---|---|---|---|---|
| Gemini 3 Flash | Validation | 1 | 74.4 | 88.8 | 14.4 | 2.73 | [2.26, 3.29] | < .001 |
| Gemini 3 Flash | Validation | 2 | 74.4 | 90.1 | 15.7 | 3.13 | [2.58, 3.81] | < .001 |
| Gemini 3 Flash | Validation | 3 | 74.4 | 89 | 14.6 | 2.78 | [2.29, 3.38] | < .001 |
| Claude Sonnet 4.6 | Validation | 1 | 81.5 | 92.7 | 11.2 | 2.88 | [2.27, 3.67] | < .001 |
| Claude Sonnet 4.6 | Validation | 2 | 81.5 | 92 | 10.5 | 2.61 | [2.07, 3.29] | < .001 |
| Claude Sonnet 4.6 | Validation | 3 | 81.5 | 93.3 | 11.8 | 3.16 | [2.44, 4.09] | < .001 |
| ChatGPT-5.3 | Validation | 1 | 96.2 | 96.9 | 0.7 | 1.23 | [0.80, 1.91] | 1 |
| ChatGPT-5.3 | Validation | 2 | 96.2 | 96.4 | 0.2 | 1.06 | [0.70, 1.60] | 1 |
| ChatGPT-5.3 | Validation | 3 | 96.2 | 96 | −0.2 | 0.95 | [0.64, 1.41] | 1 |
| DeepSeek-V4-Flash | Validation | 1 | 88.7 | 89.4 | 0.7 | 1.07 | [0.87, 1.33] | 1 |
| DeepSeek-V4-Flash | Validation | 2 | 88.7 | 89.4 | 0.7 | 1.07 | [0.87, 1.33] | 1 |
| DeepSeek-V4-Flash | Validation | 3 | 88.7 | 90.6 | 1.9 | 1.23 | [1.00, 1.51] | 0.644 |
| Gemini 3 Flash | Indirectness | 1 | 39.4 | 59.4 | 20 | 2.25 | [1.95, 2.59] | < .001 |
| Gemini 3 Flash | Indirectness | 2 | 39.4 | 61.9 | 22.5 | 2.5 | [2.17, 2.88] | < .001 |
| Gemini 3 Flash | Indirectness | 3 | 39.4 | 60.5 | 21.1 | 2.36 | [2.03, 2.73] | < .001 |
| Claude Sonnet 4.6 | Indirectness | 1 | 33.1 | 63.9 | 30.8 | 3.58 | [3.05, 4.20] | < .001 |
| Claude Sonnet 4.6 | Indirectness | 2 | 33.1 | 64 | 30.9 | 3.59 | [3.07, 4.20] | < .001 |
| Claude Sonnet 4.6 | Indirectness | 3 | 33.1 | 64.9 | 31.8 | 3.74 | [3.20, 4.37] | < .001 |
| ChatGPT-5.3 | Indirectness | 1 | 74.8 | 91.8 | 17 | 3.77 | [2.94, 4.84] | < .001 |
| ChatGPT-5.3 | Indirectness | 2 | 74.8 | 93.4 | 18.6 | 4.77 | [3.63, 6.26] | < .001 |
| ChatGPT-5.3 | Indirectness | 3 | 74.8 | 91.8 | 17 | 3.77 | [2.94, 4.84] | < .001 |
| DeepSeek-V4-Flash | Indirectness | 1 | 40.3 | 45.9 | 5.6 | 1.26 | [1.08, 1.46] | 0.031 |
| DeepSeek-V4-Flash | Indirectness | 2 | 40.3 | 45.7 | 5.4 | 1.25 | [1.07, 1.45] | 0.047 |
| DeepSeek-V4-Flash | Indirectness | 3 | 40.3 | 46 | 5.7 | 1.26 | [1.09, 1.46] | 0.024 |
| Gemini 3 Flash | Framing | 1 | 63.4 | 58.2 | −5.2 | 0.8 | [0.71, 0.91] | 0.005 |
| Gemini 3 Flash | Framing | 2 | 63.4 | 60.4 | −3.0 | 0.88 | [0.78, 1.00] | 0.56 |
| Gemini 3 Flash | Framing | 3 | 63.4 | 58.2 | −5.2 | 0.8 | [0.71, 0.91] | 0.007 |
| Claude Sonnet 4.6 | Framing | 1 | 44.9 | 60.3 | 15.4 | 1.86 | [1.64, 2.12] | < .001 |
| Claude Sonnet 4.6 | Framing | 2 | 44.9 | 58.4 | 13.5 | 1.72 | [1.52, 1.95] | < .001 |
| Claude Sonnet 4.6 | Framing | 3 | 44.9 | 61 | 16.1 | 1.92 | [1.69, 2.18] | < .001 |
| ChatGPT-5.3 | Framing | 1 | 62.8 | 92.5 | 29.7 | 7.31 | [5.78, 9.24] | < .001 |
| ChatGPT-5.3 | Framing | 2 | 62.8 | 93 | 30.2 | 7.87 | [6.16, 10.00] | < .001 |
| ChatGPT-5.3 | Framing | 3 | 62.8 | 92.7 | 29.9 | 7.52 | [5.93, 9.54] | < .001 |
| DeepSeek-V4-Flash | Framing | 1 | 62 | 61.9 | −0.1 | 1 | [0.87, 1.13] | 1 |
| DeepSeek-V4-Flash | Framing | 2 | 62 | 63.8 | 1.8 | 1.08 | [0.94, 1.24] | 1 |
| DeepSeek-V4-Flash | Framing | 3 | 62 | 63.7 | 1.7 | 1.08 | [0.94, 1.23] | 1 |
| Gemini 3 Flash | Moral sycophancy | 1 | 11.1 | 21.5 | 10.4 | 2.2 | [1.87, 2.60] | < .001 |
| Gemini 3 Flash | Moral sycophancy | 2 | 11.1 | 20.8 | 9.7 | 2.1 | [1.80, 2.46] | < .001 |
| Gemini 3 Flash | Moral sycophancy | 3 | 11.1 | 20.3 | 9.2 | 2.04 | [1.75, 2.38] | < .001 |
| Claude Sonnet 4.6 | Moral sycophancy | 1 | 22.1 | 38.1 | 16 | 2.17 | [1.92, 2.45] | < .001 |
| Claude Sonnet 4.6 | Moral sycophancy | 2 | 22.1 | 38.3 | 16.2 | 2.19 | [1.94, 2.47] | < .001 |
| Claude Sonnet 4.6 | Moral sycophancy | 3 | 22.1 | 38.5 | 16.4 | 2.21 | [1.95, 2.49] | < .001 |
| ChatGPT-5.3 | Moral sycophancy | 1 | 20.4 | 45 | 24.6 | 3.21 | [2.76, 3.73] | < .001 |
| ChatGPT-5.3 | Moral sycophancy | 2 | 20.4 | 44.9 | 24.5 | 3.18 | [2.73, 3.71] | < .001 |
| ChatGPT-5.3 | Moral sycophancy | 3 | 20.4 | 43.9 | 23.5 | 3.06 | [2.62, 3.58] | < .001 |
| DeepSeek-V4-Flash | Moral sycophancy | 1 | 82.3 | 43.3 | −39.0 | 0.16 | [0.14, 0.20] | < .001 |
| DeepSeek-V4-Flash | Moral sycophancy | 2 | 82.3 | 42.5 | −39.8 | 0.16 | [0.13, 0.19] | < .001 |
| DeepSeek-V4-Flash | Moral sycophancy | 3 | 82.3 | 43.3 | −39.0 | 0.16 | [0.14, 0.20] | < .001 |

**Table A7.** Run-specific robustness of UI–API-default differences in social sycophancy across the three UI repetitions. Δ pp = UI − API default difference in percentage points. ORs compare the indicated UI repetition with API default. Each UI repetition was analyzed separately. Within each repetition, *p* values for validation, indirectness, and framing were Bonferroni-adjusted across the 12 model-by-dimension comparisons. Moral-sycophancy *p* values were adjusted separately across the four model-specific comparisons. CIs were not adjusted for multiple comparisons.

**Prompt-sentiment robustness analysis.** We examined whether the primary UI–API-default differences were robust to adjustment for prompt sentiment. Across all four sentiment operationalizations, adjustment produced only minimal changes in the estimated access-mode effects (Table A8). The mean absolute change in the OR ranged from 0.053% to 0.159%, and the largest change observed across all 64 specifications was 1.097%. None of the adjusted analyses changed the significance status of the corresponding unadjusted comparison.

| Sentiment classifier | Comparisons, *n* | Mean absolute OR change | Maximum absolute OR change | Significance changes |
|---|---|---|---|---|
| CardiffNLP | 16 | 0.135% | 1.080% | 0 |
| XLM-R | 16 | 0.159% | 0.974% | 0 |
| Siebert | 16 | 0.053% | 0.240% | 0 |
| GoEmotions | 16 | 0.127% | 1.097% | 0 |

**Table A8.** Robustness of UI–API differences to adjustment for prompt sentiment